\documentclass[twocolumn,prl,aps,showpacs,amsmath,amssymb,floatfix]{revtex4}

\usepackage[pdftex]{graphicx}
\usepackage[pdftex]{color}

\def\lsim{\mathrel{\raise.3ex\hbox{$<$\kern-.75em\lower1ex\hbox{$\sim$}}}}
\def\gsim{\mathrel{\raise.3ex\hbox{$>$\kern-.75em\lower1ex\hbox{$\sim$}}}}

\begin{document}

\title{SNe observations in a meatball universe with a local void}

\author{Kimmo Kainulainen} \email{kimmo.kainulainen@phys.jyu.fi}
\affiliation{Department of Physics, University of Jyv\"{a}skyl\"{a}, PL 35 (YFL), 
FIN-40014 Jyv\"{a}skyl\"{a}, Finland}
\affiliation{Helsinki Institute of Physics, University of Helsinki, PL 64, 
FIN-00014 Helsinki, Finland}

\author{Valerio Marra} \email{valerio.marra@jyu.fi}
\affiliation{Department of Physics, University of Jyv\"{a}skyl\"{a}, PL 35 (YFL), 
FIN-40014 Jyv\"{a}skyl\"{a}, Finland}
\affiliation{Helsinki Institute of Physics, University of Helsinki, PL 64, 
FIN-00014 Helsinki, Finland}

\begin{abstract}
We study the impact of cosmic inhomogeneities on the interpretation of  observations.  We build an inhomogeneous universe model without dark energy that can confront supernova data and yet is reasonably well compatible with the Copernican Principle. Our model combines a relatively small local void, that gives apparent acceleration at low redshifts, with a meatball model that gives sizeable lensing (dimming) at high redshifts. Together these two elements, which focus on different effects of voids on the data, allow the model to mimic the concordance model.
\end{abstract}

\pacs{95.36.+x, 98.62.Sb, 98.65.Dx, 98.80.Es}

\maketitle

%%%%%%%%%%%%%%%%%%%%%%%%%%
%%%%%%%%%%%%%%%%%%%%%%%%%%
\paragraph{Introduction.}
%%%%%%%%%%%%%%%%%%%%%%%%%%
%%%%%%%%%%%%%%%%%%%%%%%%%%

The ``safe'' consequence of the success of the concordance model is that the flat isotropic and homogeneous $\Lambda$CDM model is a good phenomenological fit to the real inhomogeneous universe. We will discuss the possibility that the late time evidence for dark energy can be explained by the coincidental late time formation of nonlinear large scale inhomogeneities. Indeed, until the nature of the dark energy is completely understood, it is useful to look for alternative models that fit the data.

We will assume that the spacetime of the inhomogeneous universe is accurately described by small perturbations around the Friedmann-Lema\^{i}tre-Robertson-Walker (FLRW) solution whose energy content and spatial curvature are defined as Hubble-volume spatial averages over the inhomogeneous universe. Following Ref.~\cite{Kolb:2009rp} the latter will be called the Global Background Solution (GBS). The cosmological model obtained through observations, on the other hand, will be called the Phenomenological Background Solution (PBS). The idea is to associate the concordance model with the PBS, while the actual GBS is the Einstein-de Sitter (EdS) model.

\begin{figure}[h!]
\begin{center}
\includegraphics[width=7.5 cm]{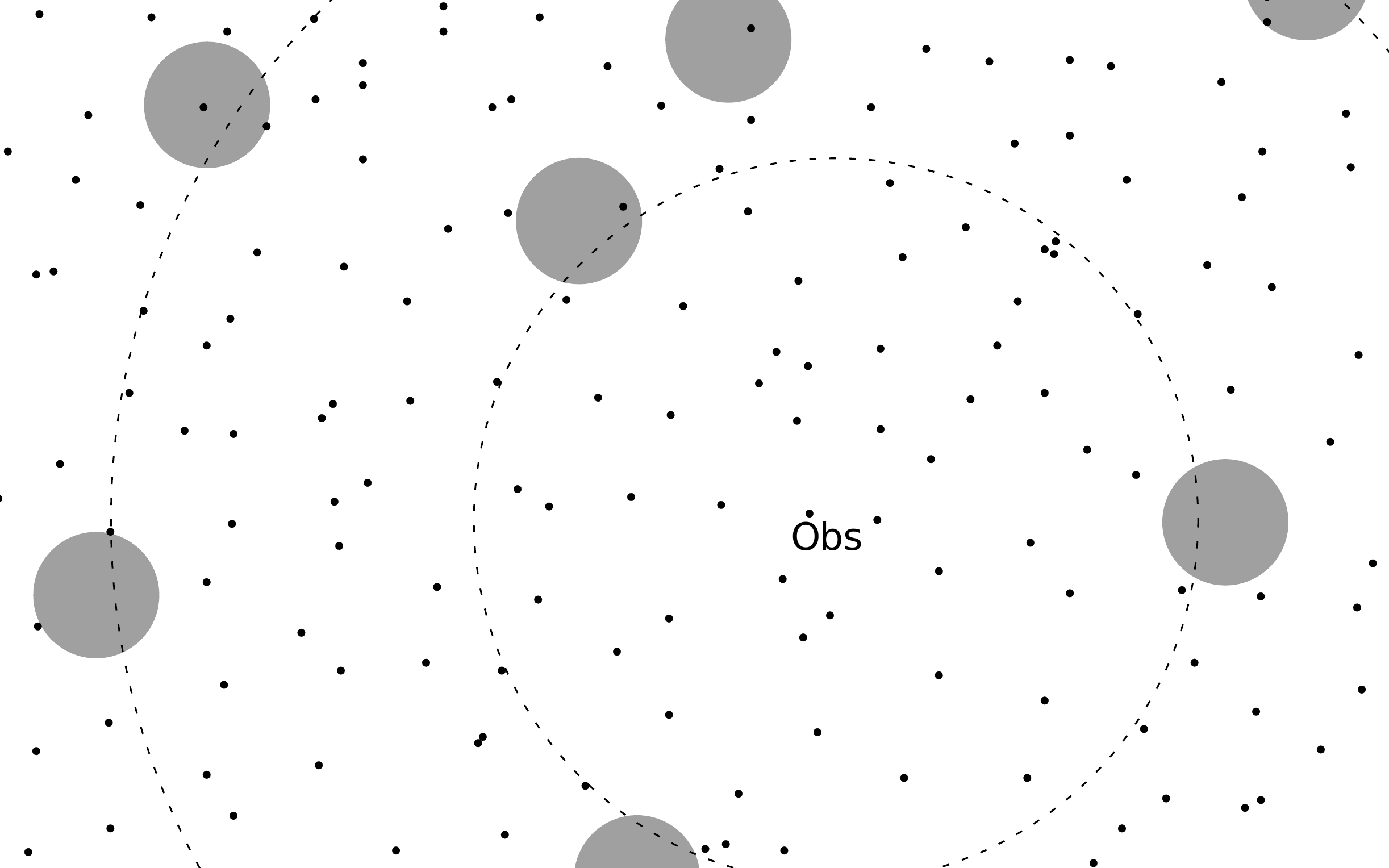}
\caption{The larger disks represent the meatballs of the family $A$, while the smaller ones represent the meatballs of the family $B$. The concentric circles mark the overdense shells in the LTB bubble and are roughly $100h^{-1}$Mpc apart.}
\label{isketch}
\end{center}
\end{figure}

%%%%%%%%%%%%%%%%%%%%%%%%%%
%%%%%%%%%%%%%%%%%%%%%%%%%%
\paragraph{Setup.}
%%%%%%%%%%%%%%%%%%%%%%%%%%
%%%%%%%%%%%%%%%%%%%%%%%%%%
For consistency with the CMB spectrum and the age of the universe,
the current background expansion rate of the EdS model is taken to be $H_{\infty}=100 \, h_{\infty}$ km s$^{-1}$ Mpc$^{-1}$ with $h_{\infty}=0.5$. The local expansion rate $H_0$ will be higher, about $h\simeq 0.58$ in the current model, which is within 2-$\sigma$ of the HST key project~\cite{Freedman:2000cf} value of $0.72 \pm 0.08$ \cite{Riess:2009pu}. Our setup is made of two elements. First, we will model the overall universe by a meatball model \cite{mbTopology} consisting of two families of randomly placed halos. The family $A$ describes very large, low density contrast structures and the family $B$ models large clusters of galaxies. The parameters specifying the model are (for each family) the average comoving distance between meatballs $d_{c}$, the proper radius of the meatball $R_p$ and the mass of the meatball $M$. The numerical values we used are given in Table \ref{param}.  The distance $d_{c}$ between meatballs is connected to their comoving number density by $n_{c}={\Gamma(4/3)^{3} \over (4 \pi/3)} d_{c}^{-3}$. We assume the meatballs to have virialized at $z=1.6$ and therefore their proper radii $R_p$ are constants. For the family $B$ we used the singular isothermal spheres (SIS) density profile and defined the radius to have a density contrast of $200$ at virialization. For the family $A$ the gaussian profile was used and the radius was chosen to qualitatively describe the largest structures seen in simulations of structure formation.
%
%%%%%%%%%%%%%%%%%%%%%%%%%%
\begin{table}
\caption{\label{param} Parameters of the meatball model ($a_0\equiv 1$).}
\begin{ruledtabular}
\begin{tabular}{lccr}
Quantity          & Family $A$    & Family $B$                        \\ 
\hline
%fraction of tot.~density & $ 1/2 $ & $1/2$              \\
$d_c$ & $100 \, h^{-1}$ Mpc         & $10 \, h^{-1}$ Mpc   \\
$R_p$  & $10 \, h^{-1}$ Mpc    & $580 \, h^{-1}$ kpc   \\
$M$    & $6.1  \cdot 10^{17} h^{-1} M_{\odot}$    & $6.1 \cdot 10^{14} h^{-1} M_{\odot}$  \\
density profile   & Gaussian    & SIS 
\end{tabular}
\end{ruledtabular}
\end{table}
%%%%%%%%%%%%%%%%%%%%%%%%%%
%
The distribution of family $A$ meatballs leaves underdense regions of order $100h^{-1}$Mpc which are filled by the family $B$ meatballs. Because the mass is equally subdivided between the two families, these underdense regions have on average a density contrast of $\delta \approx -0.5$.
The second element of our setup consists of placing the observer in one of the large underdense regions and modelling the metric of this particular void around the observer more accurately with a Lema\^{i}tre-Tolman-Bondi (LTB) bubble matched to the EdS background metric. 
See Fig.~\ref{isketch} for a sketch.
%%%%%%%%%%%%%%%%%%%%
%%%%%%%%%%%%%%%%%%%%
%%%%%%%%%%%%%%%%%%%%%
%%%%%%%%%%%%%%%%%%%%%

%%%%%%%%%%%%%%%%%%%%%%%%%%
%%%%%%%%%%%%%%%%%%%%%%%%%%
\paragraph{Lensing.}
%%%%%%%%%%%%%%%%%%%%%%%%%%
%%%%%%%%%%%%%%%%%%%%%%%%%%

In a late-time universe dominated by voids the  
homogeneity is recovered only on scales larger than the largest inhomogeneity scale, which in our model is  $L_{hom} \sim 100h^{-1}$Mpc. However, type Ia supernovae (SNe) probe angular scales $L_{SNe} \ll L_{hom}$ and it is therefore not clear if the physics inferred from SNe observations can be directly associated with the smoothed-out GBS model, especially with small data samples.
Epigrammatically, {\it the commutativity between averaging and measuring is not guaranteed} \cite{Kolb:2009rp}.
Note that this idea, which defines the weak-backreaction,
is different from the possible non-commutativity between averaging and dynamics which is the kernel of strong-backreaction studies \cite{strongbackreaction}.

Cumulative gravitational lensing is one possible source for the non-commutativity between averaging and measuring and
to study this effect we have focused on a meatball model in this letter.
This model incorporates quantitatively the crucial feature that photons can travel through voids and miss the localised overdensities.
This feature, instead, is not present in swiss-cheese models where the bubble boundaries are designed to have compensating overdensities. Such models have indeed been shown to have on average little lensing effects \cite{SCpapers,SClensing}.
In the weak-lensing approximation the lens convergence $\kappa$ is given by
\begin{equation} \label{kappa}
\kappa(z)=\int_{0}^{r_{s}(z)} dr \, G(r,r_{s}(z)) \, \delta(r,t(r)) \, ,
\end{equation}
where $\delta(r,t)$ is the density contrast and
\begin{equation}
G(r,r_{s})={3 H_{\infty}^{2} \over 2 c^{2}} {r(r_{s}-r) \over r_{s}} {1 \over a(t(r))} \, .
\end{equation}
The functions $a(t)$, $t(r)$ and $r(z)$ correspond to the EdS background model, $r_{s}$ is the comoving position of the source at redshift $z$ and the integral is evaluated along the unperturbed light path. Neglecting the second-order contribution of the shear, the shift in the distance modulus caused by lensing is
\begin{equation} \label{muu}
\Delta m (z)=5 \log_{10}(1-\kappa(z)) \, .
\end{equation}
$\Delta m$ is proportional to the total matter column density along the photon path. For a lower-than-GBS column density the light is demagnified (empty beam $\delta=-1$, for example), while in the opposite case it is magnified.

In Ref.~\cite{Kainulainen:2009dw} we derived a fast and easy way to obtain the convergence probability distribution function (PDF) for meatball models. First define a normalized profile $\varphi(x)=\rho(x)/\bar{\rho}$, where $\rho$ is the density profile of a meatball and $\bar{\rho}$ is the EdS density. Then the following quantity characterizes the contribution from one meatball, hit with an impact parameter $b$, to the convergence:
\begin{equation}
\Gamma(b,t)=\int_{b}^{R(t)} {2 \, x \, dx \over \sqrt{x^{2}-b^{2}}}\varphi(x,t)  \, .
\end{equation}
Then divide the comoving distance $r_{s}$ to the source and the radius $R$ of the meatball into $N_{S}$ and $N_{R}$ bins of widths $\Delta r_i$ and $\Delta b_m$, such that $\Delta b_m \ll R$ and $R \ll \Delta r_i \ll r_{s}$. The convergence due to a meatball placed within the bin $(i,m)$ is just
\begin{equation}
\kappa_{1im} = G( r_i,r_s) \, \Gamma ( b_m,t_i) \, .
\end{equation}
The PDF is then generated by the following functional
\begin{equation} \label{kappa3}
\kappa(\{k_{im}\}) = \sum_{i=1}^{N_S} \sum_{m=1}^{N_R}  \kappa_{1im} \left({k_{im} \over N_{O}} -\Delta N_{im} \right) \,,
\end{equation}
where $k_{im}$ is a Poisson random variable of parameter $N_{O}\Delta N_{im}$. $N_{O}(z)$ is the number of observations at redshift $z$ and $\Delta N_{im} = n_c \Delta V_{im}$ is the expected number of meatballs in the volume $\Delta V_{im} = 2\pi b_m\Delta b_m \Delta r_i$. For a Poisson variable of parameter $\lambda$, the mean is $\lambda$ and so the expected convergence in Eq.~(\ref{kappa3}) is zero, consistent with photon conservation in weak lensing. The PDF is generated by creating a large sample of configurations $\{k_{im}\}$ drawn from the Poisson distribution, and computing the appropriate convergences through Eq.~(\ref{kappa3}).
This is a task that a laptop fullfils in less than a second as opposed to expensive ray tracing techniques. The weak lensing approximation of Eq.~(\ref{kappa3}) describes the mode of the PDF very well always and, in case of non-point-like meatballs, also the tails of the PDF with $\lesssim 5 \%$ of error.

%%%%%%%%%%%%%%%%%%%%
%%%%%%%%%%%%%%%%%%%%
\begin{figure}
\begin{center}
\includegraphics[width=7.5 cm]{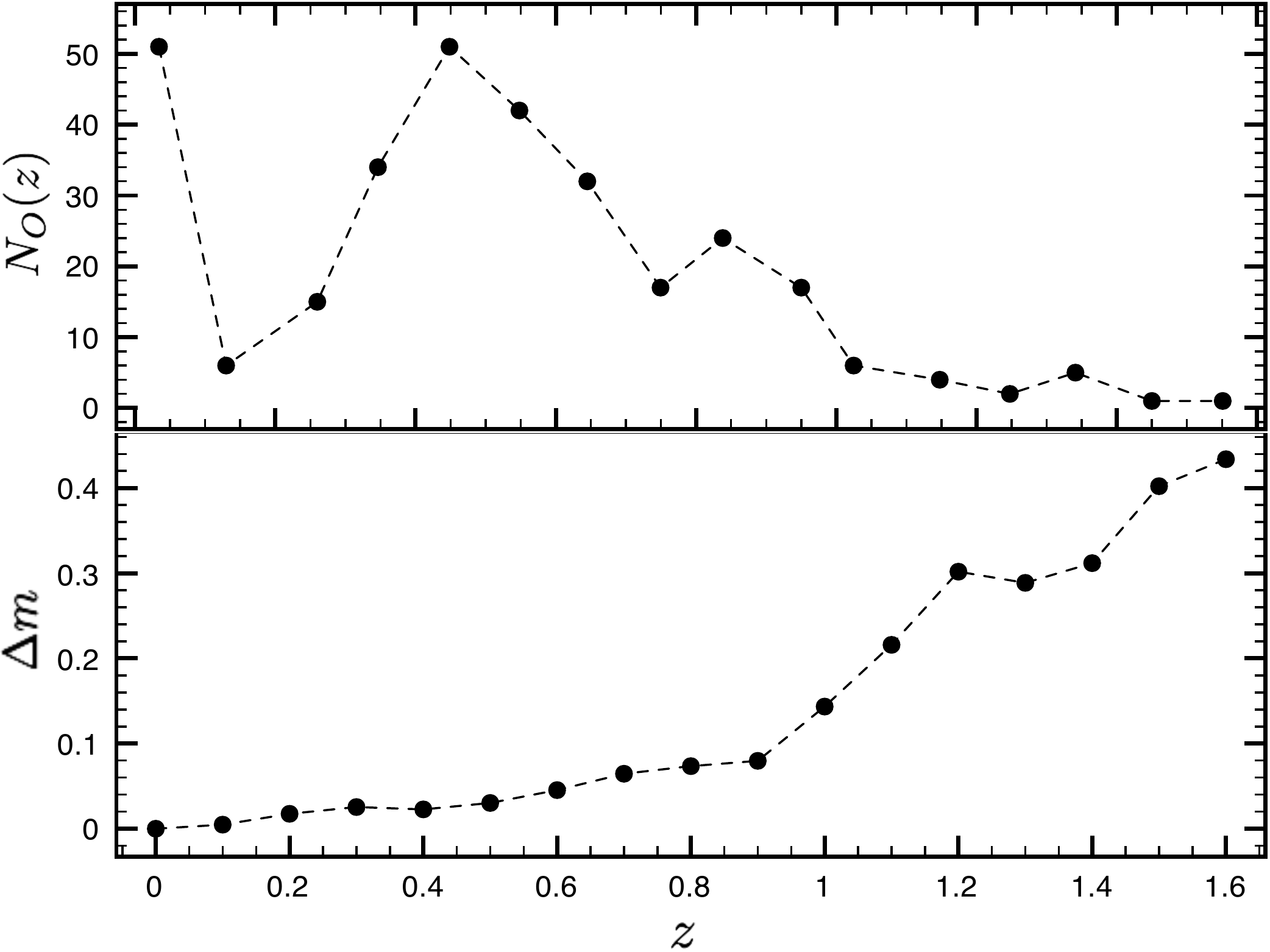}
\caption{Top: the $307$ SNe of the Union Compilation of Ref.~\cite{Kowalski:2008ez} binned with $\Delta z=0.1$.
Bottom: shift in the distance modulus $\Delta m$ for $N_O$ taken from the top panel and for the meatball model of this paper.}
\label{lensing}
\end{center}
\end{figure}
%%%%%%%%%%%%%%%%%%%%%
%%%%%%%%%%%%%%%%%%%%%

Eq.~(\ref{kappa3}) displays explicitly the effect of the size of the data sample: even if $\kappa$ might have a skewed PDF and a nonzero mode for $N_{O}=1$, the gaussianity is recovered for large $N_{O}$ and the set of observations becomes unbiased. To define $N_{O}(z)$ we will use the Union Compilation of Ref.~\cite{Kowalski:2008ez}, which consists of $307$ SNe spread between $0<z<1.6$. We have binned the SNe with a bin width of $\Delta z=0.1$ and the result is plotted in the top panel of Fig.~\ref{lensing}. SNe observations within the same bin are then be treated as repetition of the same observation.
We have used a modification of the {\tt turboGL} package \cite{turboGL} to evaluate the magnification bias for our model and for the Union Compilation. Results are shown in the bottom panel of Fig.~\ref{lensing}. The magnification bias is nonneglible for $z\gsim 0.5$ and large for $z \gsim 1$.
We stress that the latter bias is statistical and not coming from selection effects.

%%%%%%%%%%%%%%%%%%%%%%%%%%
%%%%%%%%%%%%%%%%%%%%%%%%%%
\paragraph{Hubble bubble.}
%%%%%%%%%%%%%%%%%%%%%%%%%%
%%%%%%%%%%%%%%%%%%%%%%%%%%

The second element of our model consists of using an LTB bubble matched to the EdS background of the meatball model to describe the local metric around the observer. The new ingredient is that an observer inside a void expanding faster than the background sees an apparent acceleration (see, e.g., Refs.~\cite{LTB,bigLTB}). This effect is easy to understand: our cosmological observables are confined to the light cone and hence temporal changes can be associated with spatial changes along photon geodesics. For example ``faster expansion now than before" is simply replaced by ``faster expansion here than there".  This is why a local hubble bubble model can mimic the effect of a cosmological constant at recent times ($z\lesssim 1$).
The price to pay in a simple hubble bubble model is that the inhomogeneity has to extend to the point in space/time where the effect of dark energy disappears. This requires~\cite{bigLTB} an enormous local void of radius $1.5-2 \; h^{-1}$Gpc. Moreover, to avoid a too large dipole in the CMB, our position in the void would have to be very special leading to a gross violation of the Copernican principle. The model of this letter features inhomogeneities on much smaller scales. Here the effects of the dark energy at large redshift ($z\gtrsim 0.5$) are mimicked by the statistical lensing bias, and the local void only needs to model the cosmological constant at small redshifts where the lensing effects are neglible.

%
%%%%%%%%%%%%%%%%%%%%
%%%%%%%%%%%%%%%%%%%%
\begin{figure}
\begin{center}
\includegraphics[width=7.5 cm]{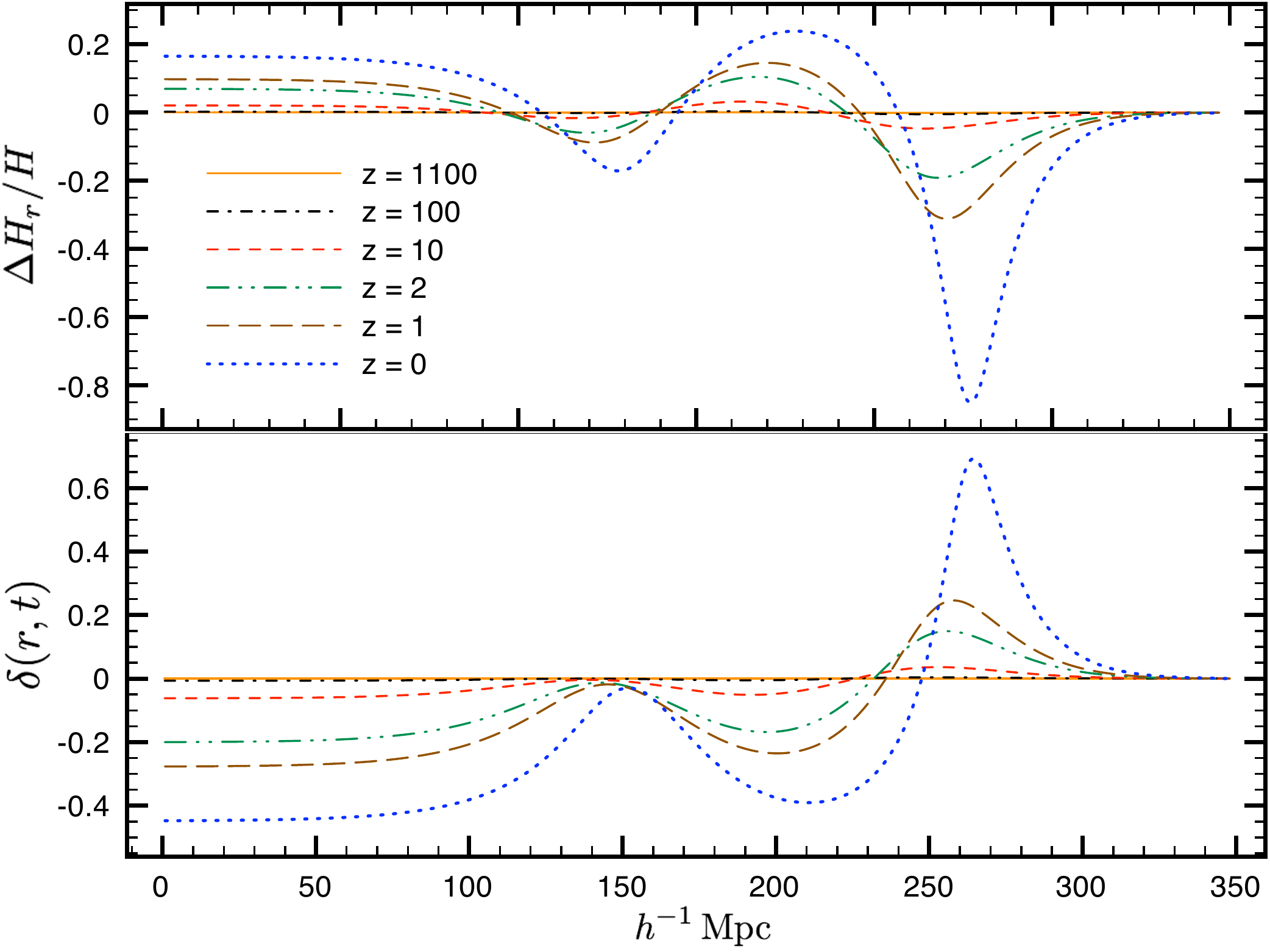}
\caption{$\Delta H_{r}/H$ (top panel) and density contrast (bottom panel) with respect to normalized proper distance ($R/a(t)$) at times corresponding to the redshifts indicated in the plot.}
\label{inico}
\end{center}
\end{figure}
%%%%%%%%%%%%%%%%%%%%%
%%%%%%%%%%%%%%%%%%%%%
%

Spherically symmetric LTB metrics can be written as
\begin{equation} \label{ltbmetric}
ds^2=-c^{2} dt^2 + \frac{R'^2(r,t)}{1-k(r)r^2} dr^2 + R^2(r,t) \, d\Omega^2 \, .
\end{equation}
This reduces to the usual FLRW metric when $R(r,t)/r \equiv a(r,t) \rightarrow a(t)$ and $k(r) \rightarrow \pm1$ or $0$.
In the dust case the Einstein equations for this metric give
\begin{equation} \label{Hb}
\frac{\dot{a}^{2}(r,t)}{a^{2}(r,t)} = \frac{8 \pi G}{3}\, 
\hat{\rho}(r,t)  - \frac{c^{2}k(r)}{a^{2}(r,t)} ,
\end{equation}
where $\hat{\rho}$ is the average density up to the shell $r$. 
We specify our LTB model at the recombination time with a density contrast of order $\sim 10^{-3}$ in accordance with the CMB.
In particular, the curvature function is defined by 2 overall parameters that specify size and depth of the local void and by 3 parameters that specify the detailed profile inside the void which is shown by the evolution of the density contrast in Fig.~\ref{inico}. Also shown is the ratio $\Delta H_{r}/H$  where $\Delta H_{r}=H_{r}-H$, $H$ is the EdS expansion rate and $H_{r}=\dot{R}'/R'$.
%is the radial expansion rate in the LTB bubble.
At late times the initial perturbations generate two fast expanding ($\Delta H_{r}>0$) voids of radius $50-100 \, h^{-1}$Mpc, surrounded by two thin collapsing ($\Delta H_{r}<0$) shells. The inner void of radius $\sim 100 \, h^{-1}$Mpc, where the observer is located, has a local expansion rate of
%$H_{0}=100 \, h$ km s$^{-1}$ Mpc$^{-1}$ with
$h \simeq0.58$ which, as anticipated, is larger than the EdS value.
Moreover, the density contrast of the latter is close to $\delta \approx -0.5$ consistently with our meatball model for the overall universe.
In order not to have a too large dipole in the CMB the observer has to be at a distance $\le 15\, h^{-1}$Mpc from the center of the inner void. Tighter constrains will come from future probes sensitive to off-center anisotropy (see, for example, \cite{Quartin:2009xr}).

%%%%%%%%%%%%%%%%%%%%%%%%%%
%%%%%%%%%%%%%%%%%%%%%%%%%%
\paragraph{Results.}
%%%%%%%%%%%%%%%%%%%%%%%%%%
%%%%%%%%%%%%%%%%%%%%%%%%%%

%
%%%%%%%%%%%%%%%%%%%%
%%%%%%%%%%%%%%%%%%%%
\begin{figure}
\begin{center}
\includegraphics[width=7.5 cm]{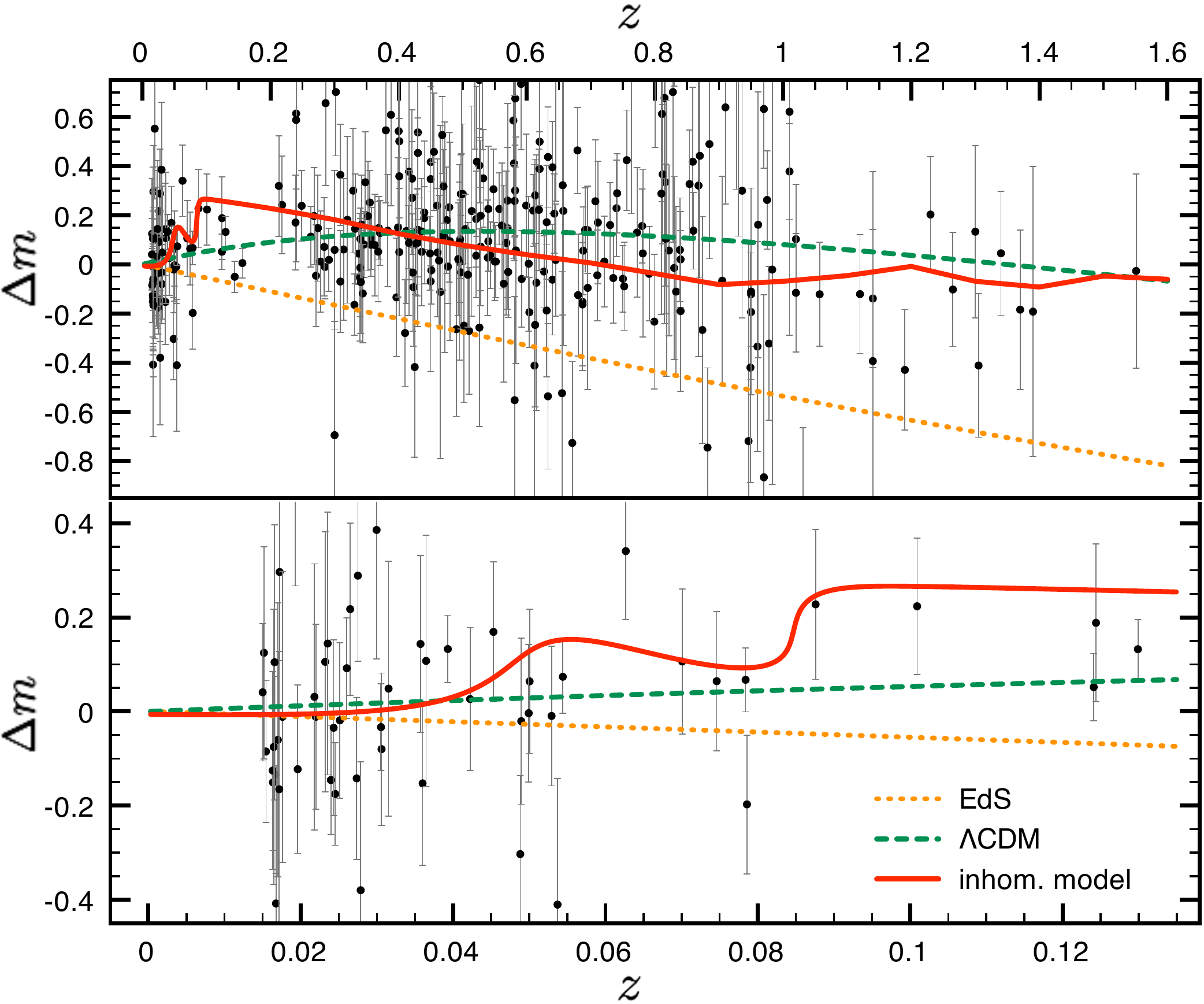}
\caption{Top panel: shown is the distance modulus with respect to the empty universe for the inhomogeneous model of this paper, for the $\Lambda$CDM model and for the EdS model together with the full Union Compilation of Ref.~\cite{Kowalski:2008ez}. Bottom panel, zoom for low redshifts.}
\label{lumi}
\end{center}
\end{figure}
%%%%%%%%%%%%%%%%%%%%%
%%%%%%%%%%%%%%%%%%%%%
%

Let us now describe the main features of our results shown in Fig.~\ref{lumi}.
First, since our observer is inside a negative-curvature dominated region, the slope of the inhomogeneous model starts almost flat at $z=0$, as it is evident in the zoom of the bottom panel. However, at finite redshifts the spatial variation of the expansion rate mimics the apparent acceleration and a positive $\Delta m$ proportional to the difference between local and background expansion rates $H_{0}-H_{\infty}$ appears. Beyond the dimensions of the local hubble bubble, at $z\gsim 0.1$, the slope of the curve follows the EdS model until the lensing bias effect becomes important at $z\gsim 0.5$ after which it starts to pull the curve up again as shown in the top panel of fig.~\ref{lumi}. This 
smooth rise of the predicted $\Delta m$ at large $z$ and the behaviour around $z\sim 0.1-0.2$, also clearly visible in Figure~\ref{lumi}, are the main characteristic differences between our model and the $\Lambda$CDM model. Both these issues would be easily resolved by a JDEM-like survey with a data set of $2000$ SNe.

We also computed the $\chi^{2}$ in order to estimate the goodness-of-fit of our  model. It should be stressed that we did not explore the parameter space to minimize the $\chi^{2}$ but chose the parameters essentially by hand. Thus the comparision between our model and the concordance model is qualitative only. Numerical results are shown in Table~\ref{chi2}. Obviously the simple EdS model has a poor fit, but introducing inhomogeneities improves the $\chi^{2}$ significantly. The local Hubble bubble has a larger impact on $\chi^{2}$, but lensing also has a large positive effect.
%
%%%%%%%%%%%%%%%%%%%%%%%%%%
\begin{table}[h!]
\caption{\label{chi2} $\chi^{2}$ for Union Compilation of $307$ SNe.}
\begin{ruledtabular}
\begin{tabular}{lccr}
Model          & $\chi^{2}$                       \\ 
\hline
$\Lambda$CDM   & 312     \\
$\Lambda$CDM + meatballs  & 323     \\
EdS        & 608   \\
EdS + H.~bubble      & 440     \\
EdS + H.~bubble  + meatballs   & 396   \\
\end{tabular}
\end{ruledtabular}
\end{table}
%%%%%%%%%%%%%%%%%%%%%%%%%%

Finally, including inhomogeneities changes also the predictions of the $\Lambda$CDM model. The second line in Table~\ref{chi2} shows results of a simulation in the context of a $\Lambda$CDM model with global parameters given in Ref.~\cite{Kowalski:2008ez}, with the same meatball mass spectrum which we used with the EdS model. Interestingly, including inhomogeneities to the concordance model makes the fit worse and suggests, as discussed in Ref.~\cite{Clifton:2009bp}, a smaller value of $\Omega_{\Lambda}$. Our model remains worse than the $\Lambda$CDM model even afer inclusion of inhomogeneities, but not nearly as dramatically so as the homogenous EdS model.

Of course our setup is but a simple toy model. However, we believe that our results make it explicitly clear that different effects of inhomogeneities can pull into the same direction, making the PBS depart from the GBS. Moreover, we believe that the two effects of the voids considered here are not the end of the story; to conclude on the viability of inhomogeneous universe models as the possible explanation of the apparent acceleration, it is crucial to consider all possible effects in as realistic a model as possible.

%%%%%%%%%%%%%%%%%%%%%%%%%%
%%%%%%%%%%%%%%%%%%%%%%%%%%

%%%%%%%%%%%%%%%%%%%%%%%%%%
%%%%%%%%%%%%%%%%%%%%%%%%%%

%%%%%%%%%%%%%%%%%%%%%%%%%%
%%%%%%%%%%%%%%%%%%%%%%%%%%
%%%%%%%%%%%%%%%%%%%%%%%%%%
\end{document}